\documentclass[prl,twocolumn,aps,amssymb,footinbib,showpacs]{revtex4}
\usepackage{amssymb}

\usepackage{graphicx}
\usepackage{amsmath}
\usepackage{times}

\begin{document}

\title{Metal-Insulator Transition Revisited for Cold Atoms in Non-Abelian Gauge Potentials }

\author{Indubala I. Satija$^{1,2}$, Daniel C. Dakin$^{2}$ and Charles W. Clark$^{2}$}
\email{isatija@physics.gmu.edu}
\email{daniel.dakin@nist.gov}
\affiliation{$^{1}$ Dept. of Phys., George Mason U., Fairfax, VA, 22030,USA}
\affiliation{ $^{2}$ National Institute of Standards and
Technology, Gaithersburg MD, 20899, USA}

\date{\today}
\begin{abstract}
We discuss the possibility of realizing metal-insulator transitions
with ultracold atoms in two-dimensional optical lattices in the presence of
artificial gauge potentials. Such transitions have been extensively studied for magnetic fields corresponding
to Abelian gauges;
they occur when the magnetic flux penetrating
the lattice plaquette is an irrational multiple of the magnetic flux quantum.
Here we present the first study of these transitions for non-Abelian $U(2)$ gauge fields,
which can be realized with atoms with two pairs of {\it degenerate} internal states.
In contrast to the Abelian case,
the spectrum and localization transition in the non-Abelian case is strongly influenced by atomic momenta.
In addition to determining the localization boundary, the momentum
fragments the spectrum and the
minimum energy viewed as a function of momentum exhibits a step structure.
Other key characteristics
of the non-Abelian case include the absence of localization for certain states and satellite fringes
around the Bragg peaks in the momentum distribution and an interesting possibility that the transition can be tuned
by the atomic momenta.

\end{abstract}
\maketitle

Ultracold atom research presents many avenues to explore problems at the forefront of theoretical
physics. One recent line of investigation is the generation of artificial
``magnetic'' fields in an electrically neutral quantum gas~\cite{NJP,Muller,FQHE}, which have the same effect on neutral atoms as a magnetic field has on charged particles.
For atoms confined in an optical lattice, these synthetic gauge fields can be realized by imposing appropriate complex phase factors upon the
amplitudes for tunneling of atoms between neighboring sites.
This has renewed interest in the substantial theoretical literature of diamagnetism of
electrons in incommensurate two-dimensional lattices~\cite{Harper}, a system which exhibits rich phase structure,
but has seemed difficult to realize with any clarity in a condensed-matter system.
Atoms in optical lattices may now provide a vehicle for experimental exploration of this subject.

Furthermore, it also seems possible to create {\it non-Abelian} gauge fields in optical lattice systems~\cite{NA},
fields in which the components of the vector potential are noncommuting operators rather than scalar functions.
Such fields are of fundamental importance in condensed matter and particle physics~\cite{GPbook},
and ultracold atomic physics now offers the prospect of controlled, well-characterized experiments on
them~\cite{USB}: for example, the observation of analogue ``magnetic monopoles''~\cite{Mono}
and non-Abelian interferometry~\cite{NA}.
Recently, it was suggested that the artificial magnetic fields may provide a laboratory
realization of the fractal energy spectrum
of an electron gas in a two-dimensional lattice, the Hofstadter butterfly~\cite{Hof}, and the ``Hofstadter Moth'', the corresponding spectrum
for the $U(2)$ gauge fields~\cite{NJP,NA}. Another interesting feature that can be realized in these systems
is the metal-insulator transition
 when the magnetic flux $\alpha$ per lattice plaquette is an irrational multiple of the magnetic flux quantum~\cite{Harper}.
In contrast to the Mott transition, which has its origin in particle interactions~\cite{Greiner}, this phase transitions is
induced by competing periodicities associated with the cyclotron frequency and the frequency of motion of a particle
in a periodic lattice~\cite{Hof}.

The localization transition obtained by tuning the ratio of the tunneling along the two directions
of the lattice has been extensively studied~\cite{Harper}. Here, we revisit the transition for
the case when the ``magnetic field'' corresponds to $U(2)$ gauge fields.
We describe various novelties associated with the transition in the non-Abelian case and
discuss the experimental feasibility of seeing these transitions in two-dimensional optical lattices.
To best of our knowledge, this is the first study to examine the role of gauge groups
on a quantum phase transition.

Our starting point is a tight binding model (TBM) of a particle moving on a
two-dimensional rectangular lattice $(x,y)$, with lattice constants $(a,b)$,
and nearest-neighbor hopping characterized by the tunneling amplitudes $(J, J\Lambda)$.
When a weak external vector potential,  $\vec{A}(x,y) = (A_x,A_y,0)$, is applied to the system, the Hamiltonian is given by
\begin{equation*}
H= -J\left[\cos\left((p_x-\frac{e}{c}A_x)\frac{a}{\hbar}\right)+ \Lambda \cos\left((p_y-\frac{e}{c}A_y) \frac{b}{\hbar}\right)\right],
\end{equation*}
where $\vec{p}$ is the momentum operator.
For simplicity, we work in the Landau gauge: $\vec{A}(x,y) = (A_x,A_y(x),0)$, where $A_x$ is a constant
and $A_y$ depends only on the $x$-coordinate. The resulting Hamiltonian is cyclic in $y$
and thus the eigenfunctions are proportional to plane waves in the $y$-direction.
Denoting the transverse wave number of the plane wave
as $\tilde{k}_y=k_y/b$, the wave function can be written as: $\Psi(ma,nb)=e^{ik_y n} g_m$ with $x=ma$ and $y=nb$.
For the Abelian gauge case, with $\vec{A}=(0, By, 0)$,
one obtains the Harper equation~\cite{Harper}
\begin{equation}
g_{m+1}+g_{m-1}+2\Lambda \cos(2\pi\alpha m - k_y)g_m=E\,g_m,
\label{Harp}
\end{equation}
where $\alpha=Bab/(hc/e)$ and $E(\alpha)$ is the energy in units of $J$.
The energy spectrum is the union over $k_y$ of
individual energy spectra of the Harper equation. For rational $\alpha=p/q$, the spectrum consists of $q$
bands which are usually separated by gaps. As $k_y$ varies, the bands shift and their length may change,
but they do not overlap, except at the band edges. For irrational
$\alpha$, the spectrum is independent of $k_y$ and the system exhibits a metal-insulator transition at $\Lambda=1$.
As we discuss below, this insensitivity of the spectrum and the localization transition to the transverse wave number
is lost when the gauge potential is non-Abelian.

We obtain a non-Abelian vector potential for ultracold atoms in quite a natural way, when we treat systems in
which the atoms have two degenerate internal states.  The most general vector potential will couple the two states,
and thus gives rise to $U(2)$ gauge symmetry. In our treatment of the non-Abelian case,
we follow the convention of an earlier study~\cite{NA} and adopt its standard form of the vector potential,
\begin{eqnarray*}
\vec{A}=\frac{\hbar c}{ea}\left(\frac{\pi}{2}\left( \begin{array}{cc} -1 & 1\\ 1& -1\\ \end{array}\right), \:
\left(\begin{array}{cc} 2\pi m \alpha_1 & 0\\ 0 & 2\pi m \alpha_2\\ \end{array} \right ), \: 0 \right ).
\end{eqnarray*}
Here $\alpha_i$ determines the ``magnetic flux'' of the lattice.

The wave function $\Psi(m,n)= e^{ik_yn} {\bf g}_m$ is a two-component vector
${\bf g}_m=\left (\begin{array}{cc} \theta_m\\ \eta_m \end{array} \right )$ that satisfies the following equations,
\begin{eqnarray}
\left (\begin{array}{cc} \theta_{m+1}\\ \eta_{m+1} \end{array} \right )+
\left (\begin{array}{cc} \theta_{m-1}\\ \eta_{m-1} \end{array} \right )+
\left (\begin{array}{cc} 0 &\epsilon_m^{\alpha_2}\\ \epsilon_m^{\alpha_1} & 0 \end{array} \right )
\left (\begin{array}{cc} \theta_m\\ \eta_m \end{array} \right )=0,
\label{tbmc}
\end{eqnarray}
where $\epsilon_m^{\alpha_i}=E-2\Lambda \cos(2\pi\alpha_i m-k_y)$.

These coupled equations can be decoupled resulting in a pair of TBMs,
\begin{subequations}
\begin{eqnarray}
\epsilon_{m-1}^{\alpha_2}\theta_{m+2}+\epsilon_{m+1}^{\alpha_2}\theta_{m-2}&=&C_m(\alpha_1,\alpha_2)\theta_m,\label{tbmucA}\\
\epsilon_{m-1}^{\alpha_1}\eta_{m+2}+\epsilon_{m+1}^{\alpha_1}\eta_{m-2}&=&C_m(\alpha_2,\alpha_1)\eta_m,\label{tbmucB}
\end{eqnarray}
\end{subequations}
where $C_m(\alpha_1,\alpha_2)=\epsilon_m^{\alpha_1} \epsilon_{m+1}^{\alpha_2}\epsilon_{m-1}^{\alpha_2}
-\epsilon_{m-1}^{\alpha_2}-\epsilon_{m+1}^{\alpha_2}$.

For $\alpha_1=\alpha_2$, the non-Abelian problem reduces to the Abelian
problem described by the Harper equation~[Eq.~(\ref{Harp})].
Calculation of the spectral properties and the detailed exploration of their localization transition in
the non-Abelian problem is done using various methods which include a transfer matrix approach~\cite{Hof,NA},
mapping to a dynamical system~\cite{KS} and the scaling arguments from Fibonacci renormalization theory~\cite{KSrg}.
In our study we fix $\alpha_1$ to be the golden mean, $\alpha_1=(\sqrt{5}-1)/2$;
in the transfer matrix analysis, $\alpha_1$ is approximated
by a sequence of Fibonacci numbers $F_n$: $F_1=1$, $F_2=1$, $F_{n+1}=F_n+F_{n-1}$.
The non-Abelian problem is studied for various values of $\alpha_2$ and
we describe our results for $\alpha_2=\frac{1}{2}, \frac{1}{4}$ and $\alpha_1^4$.

For the uncoupled Eq.~(\ref{tbmucA} (\ref{tbmucB})), the transfer matrix method is used to obtain the allowed
energies of the system~\cite{Harper,NA}. In general, Eqs.~(\ref{tbmucA} (\ref{tbmucB}))
give two sets of allowed energies depending upon whether $m$ is even or odd and
therefore, the full spectrum of the non-Abelian system described by Eq.~(\ref{tbmc}) is obtained by
the {\it intersection} of the two spectra obtained for the even and odd-site $\theta$ (or $\eta$) equation.
However, when $\alpha_1$ and $\alpha_2$ are both irrationals, or if one of them is zero, the spectra
of the even and odd-site equations are identical.

Localization properties, particularly the phase diagram of the system can be best studied by
mapping the TBMs (Eq.~(\ref{tbmucA}))
to a dynamical system, a discrete driven map obtained by defining
$x_m=\frac{\theta_{m-2}}{\theta_m}$~\cite{KSrg}. The $\theta$ equation becomes
$x_{m+2}=-1/[R_m x_m-f_m]$. Here $R_m=\epsilon^{\alpha_2}_{m+1}/\epsilon^{\alpha_2}_{m-1}$
and $f_m=C_m(\alpha_1,\alpha_2)/\epsilon^{\alpha_2}_{m-1}$.
This mapping of a TBM to a dynamical map is extremely useful in determining the localization
transition as the absolute value of the Lyapunov
exponent $\gamma$ is related to the inverse localization length $\xi$ of the TBM, $\xi^{-1} = \gamma/2$.
It should be noted that the maps are non-chaotic and hence the Lyapunov exponents are always negative.

\begin{figure}[htbp]
\includegraphics[width =0.9\linewidth,height=0.7\linewidth]{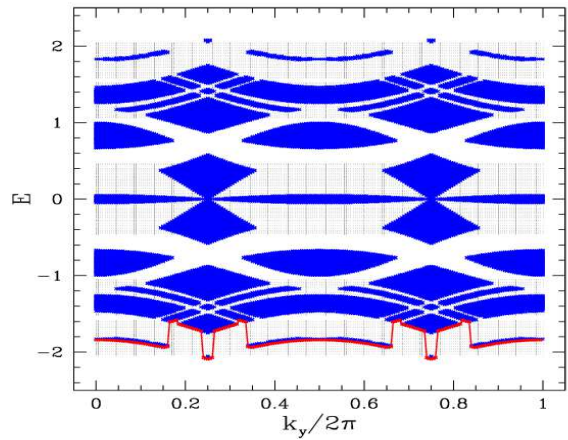}
\leavevmode \caption{(color online) Energy spectrum viewed as a function of $k_y$ for $\alpha_1=\frac{89}{144}$,
$\alpha_2=\frac{1}{2}$
and $\Lambda=0.3$. Light dots show the corresponding Abelian case for $\alpha_1=\alpha_2=89/144$ where spectrum
does not depend upon $k_y$. Minimum energy shown by a solid line shows the step structure with $\frac{k_y}{2\pi}=\frac{1}{4}$
being the global minimum or the ground state.}
\label{fig1}
\end{figure}

\begin{figure}[htbp]
\includegraphics[width =0.9\linewidth,height=0.6\linewidth]{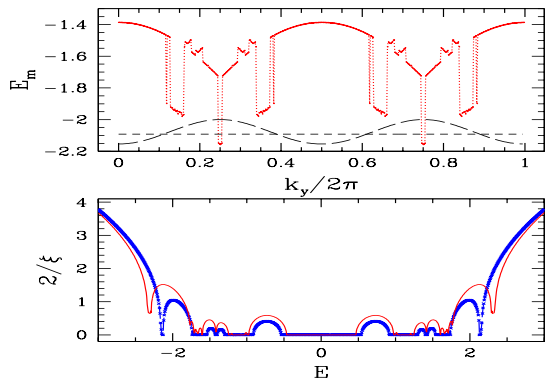}
\leavevmode \caption{(color online) Upper panel shows minimum energy for $\alpha_2=\frac{1}{2}$ with $\Lambda=0.4$.
Short and long dashed lines show the corresponding Abelian case with $\alpha =\frac{89}{144}$ and $\frac{1}{2}$,
respectively.
Lower panel illustrates the localization of the ground state ($\frac{k_y}{2\pi}=\frac{1}{4})$;
localization length $\xi$ (in units of $a$)
that with $\Lambda=0.4$ (solid line) is $\infty$ becomes finite at $\Lambda=0.6$ (dashed line).
Here $E$ is treated as a parameter: the extended states correspond to
$1/\xi \rightarrow 0$, spikes describe localized states and the smooth lobes correspond to
forbidden regions.}
\label{fig2}
\end{figure}

In contrast to the Abelian case,
the spectrum and the localization properties of states in the non-Abelian case are
strongly dependent on the transverse momentum $k_y$.
Figure~\ref{fig1} shows the fragmented energy spectrum of the $\alpha_2=\frac{1}{2}$ case as the
transverse wave vector is varied. The $k_y/2\pi=\frac{1}{4}$ emerges as a special wave number
which characterizes the ground state and is also important for the central state ($E=0$)
as three different bands touch at $E=0, \frac{k_y}{2\pi}=\frac{1}{4}$.\cite{foot}
Furthermore, the minimum energy develops a step structure
and the number of steps increases with $\Lambda$ (Fig.~\ref{fig2}) and this structure resembles a fractal set at the onset to localization.
The origin of these discontinuities is two-fold: firstly,
only a fraction of the full range of $k_y$ results in the allowed energies
and secondly, the incompatibility of the two sets of energy solutions
corresponding to the even and the odd site equations.
However, there exists a characteristic value of the wave number for which
even-odd site equations result
in identical set of allowed energies. For $\alpha_2=\frac{1}{2}$, this selected set of wave numbers is
$\frac{k_y}{2\pi}=\frac{1}{4}$ (mod $1$).
For $\alpha_2=\frac{1}{4}$, which exhibits analogous step-structure, the characteristic wave numbers are
given by $\frac{k_y}{2\pi}=\frac{1}{8}$ (mod $\frac{1}{2}$).

Figure~\ref{fig2} shows the localization aspect of the spectrum. In stark contrast to the Harper equation where
all states localize simultaneously at $\Lambda=1$, localization in the non-Abelian case begins at the band edges
at $\Lambda \approx .48$
for $\alpha_2=\frac{1}{2}$. In this case, all states except those in the central band localize before or at
$\Lambda=1$. As $\Lambda$ increases, the width of the central band decreases and
as $\Lambda \rightarrow \infty$, only $E(k_y=\pi/2)=0$
remains extended, a fact that
can be inferred from Eq.~(\ref{tbmc}) since
it corresponds to $\epsilon_m^{\alpha_2}=0$.
This defiance of localization of the central band
is a unique feature of the $\alpha_2=\frac{1}{2}$ case. For $\alpha_2=\frac{1}{4}$, localization begins
at $\Lambda \approx 0.24$ and all states localize before $\Lambda \lesssim 0.83$.

\begin{figure}[htbp]
\includegraphics[width =0.9\linewidth,height=0.5\linewidth]{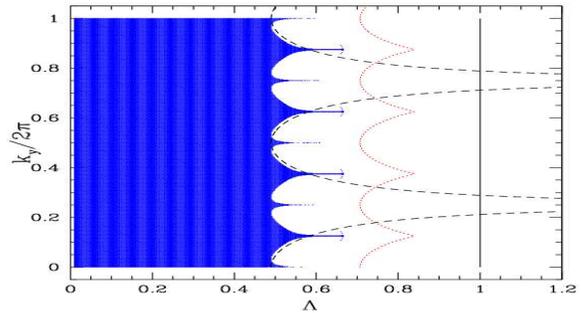}
\leavevmode \caption{(color online)
Shaded region shows the extended phase for the $E=0$ state for $\alpha_2=\alpha_1^4$.
Long and short lines show the boundary of the extended phase
for $\alpha_2=\frac{1}{2}$ and $\alpha_2=\frac{1}{4}$, respectively. Solid vertical line shows the localization boundary for the Abelian problem described by Eq.~(\ref{Harp}) with $\alpha$=golden mean.}
\label{fig3}
\end{figure}

To understand the origin of the selected set of wave numbers, we now focus on the TBM corresponding to the $E=0$ state,
which is always an eigenvalue of Eq.~(\ref{tbmc}) for irrational $\alpha_1$.
For $\alpha_2=\frac{1}{2}$, the $\theta$ equation for $E=0$ reduces to,
\begin{equation}
\theta_{m+2}+\theta_{m-2}+2\Lambda_{eff} \cos(2\pi\alpha_1 m -k_y)\theta_m= (-2)\theta_m,
\label{Hlike}
\end{equation}
where $\Lambda_{eff}=2\Lambda^2 \cos(k_y)$. Thus, the $E=0$ state of the non-Abelian system is mapped to the $E=-2$ state of a Harper-like equation
(\ref{Hlike}) where the strength of the onsite quasiperiodic potential is a sinusoidal function of $k_y$.
Diagonalization of Eq.~(\ref{Hlike})
for various values of $\Lambda_{eff}$ shows that $E=-2$ is an eigenvalue of the system
provided $\Lambda_{eff} \lesssim 0.48$.
Thus the boundary curve $E(k_y, \Lambda)=0$ for the delocalized phase of the $E=0$ state is given
by $2\Lambda^2 \cos(k_y) \approx 0.48$.
This result is in perfect agreement with the numerically obtained boundary curve in Fig.~\ref{fig3}
using dynamical mapping and renormalization methods.

A very direct evidence of the importance of the transverse wave number in the localization transition emerges for
$\alpha_2=\frac{1}{4}$. The $\theta$ equation for the $E=0$ state can be mapped to a \underline{non-hermitian} TBM,
\begin{equation}
\bar{\theta}_{m+2}+\bar{\theta}_{m-2}+2 i \Lambda_{e} \cos( 2\pi\alpha_1 m-k_y)\bar{\theta}_m=0,
\end{equation}
where $\Lambda_e=2\Lambda^2 \cos(k_y)$ if $m$ is odd and
$\Lambda_e=2\Lambda^2 \sin(k_y)$ if $m$ is even and $\bar{\theta}_m=i^m \theta_m$.
Analogous to the corresponding hermitian problem, the system exhibits self-duality at $\Lambda_e=1$ and
this self-dual point describes the onset of localization~\cite{Amin}.
Therefore, the boundary curve describing the localization transition for the $E=0$ state corresponds to the minimum value of $\Lambda$
consistent with the solutions of $2\Lambda^2 \cos(k_y)=1$ and
$2\Lambda^2 \sin(k_y)=1$. It should be noted that for $k_y/2\pi=\frac{1}{8}$~(mod $\frac{1}{2}$), even-odd boundary curves are
identical and we see the emergence of a characteristic wave number that corresponds to
the maximum value of $\Lambda$ for localization of
$E=0$ state.
These analytic predictions are
in perfect agreement with the numerical computation of the localization boundary (See Fig.~\ref{fig3}).

The above mapping elucidates
the importance of the wave number $k_y$ in localization transition in
the non-Abelian case. The transition threshold is a sinusoidal function of $k_y$ and there exists a special set of wave numbers that survive
localization the longest as we increase the parameter $\Lambda$. The importance of $k_y$ and the existence of
characteristic wave numbers  illustrated explicitly
for $E=0$ state is also valid for ground states as
it is precisely at these wave numbers that
the full energy spectrum for even and odd sites for $\theta$~(or $\eta$) are fully compatible.
This special set of wave numbers also
corresponds to the ground state of the system. Interestingly, these features also characterize the localization transition for irrational $\alpha_2$.
The localization boundary for $\alpha_2=\alpha_1^4$ and $E=0$ (See Fig.~\ref{fig3}) clearly shows the emergence of a selected set of wave numbers
$k_y/2\pi=\frac{n}{8}$~($n=0,1,2...$) that remain extended for the maximum range of $\Lambda$.
Our detailed analysis of this case shows that the localization begins at $\Lambda \approx 0.14$ and all states are localized
before $\Lambda \approx 0.7$.

We now turn our attention to the experimental aspects of creating and observing the effects
of non-Abelian potentials. A proposed experimental implementation of artificial Abelian~\cite{NJP} and non-Abelian fields~\cite{NA}
consists of a two-dimensional optical lattice populated with cold atoms that occupy two hyperfine states.
The typical kinetic energy tunneling along the $x$-direction is suppressed by accelerating the system or
applying an inhomogeneous electric field in that direction.
Tunneling is accomplished instead with laser driven Raman transitions,
induced by additional running wave lasers detuned to cancel the effect of the acceleration.
To generate non-Abelian fields, we have two sets of Raman transitions.


\begin{figure}[htbp]
\includegraphics[width =0.9\linewidth,height=0.5\linewidth]{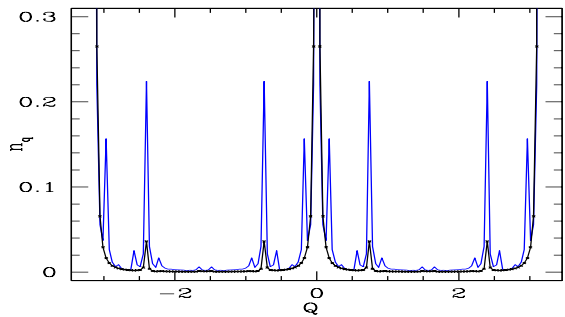}
\leavevmode \caption{(color online) Momentum distribution
for the ground state ($k_y/2\pi=\frac{1}{4}$) before it localizes for
$\alpha_1=\frac{89}{144}$ and $\alpha_2=\frac{1}{2}$ ($\Lambda=0.4$).
The line with crosses shows the corresponding result for the Abelian case.
Note that the central peak
is truncated by a factor of three to show smaller peaks.}\label{fig4}
\end{figure}

We can experimentally access various $\Lambda$ values (ratio of the tunneling in $y$ and $x$ directions)
by adjusting the lattice beam intensity.
Explicit formulas for tunneling due to kinetic energy ($J$)~\cite{Ana} and
numerical calculations of laser-induced tunneling have been reported~\cite{NJP}.
Based on those results, one can write $\Lambda/E_R = \left(1/\hbar\Omega\right) f(V_0/E_R,\alpha)$, where $f$ is a numerically known function.
Here $V_0$ is the optical potential of the lattice, $E_R$ is the lattice recoil energy, and $\Omega$ is the Rabi
frequency associated with laser-induced tunneling. In the non-Abelian case, there are generally
two possible values of $\Lambda$ corresponding to $\Omega_1$ and $\Omega_2$.
By adjusting $\Omega_2/\Omega_1=f(V_0/E_R,\alpha_1)/f(V_0/E_R,\alpha_2)$, we obtain a single $\Lambda$
in correspondence with the theoretical studies described here.
For a practical range of $V_0/E_R$ between 5 and 45, $f$ decreases monotonically from 0.5 to nearly zero.
To create a useful parameter range to see metal-insulator transition ($0 < \Lambda \lesssim 2$), we require that
the parameter $\hbar\Omega/E_R$ be set to order unity. It is possible to achieve this with
reasonable experimental settings for alkali atoms like $^{87}$Rb if the two states
are taken to be the hyperfine levels of the $5^2S_{1/2}$ state.
To begin we note that the above scheme requires the Rabi frequency of the Raman transition $\Omega$,
the detuning $\Delta$, and the lattice trapping frequency $\nu_x$ to have well separated magnitudes
such that $\Omega\ll\Delta\ll\nu_x$ to ensure that only the lowest band of the lattice is
occupied and no other excitations occur.
Typical values of $\nu_x=E_R/h$ are on the order of tens of kilohertz~\cite{Greiner}.
The Raman transition is stimulated by two lasers with Rabi frequencies $\Omega_g^{(1)}$ and $\Omega_e^{(1)}$
and a large detuning $\delta_r$ such that the effective Rabi frequency magnitude is
$\Omega=\Omega_g^{(1)}\Omega_e^{(1)}/2\delta_r$.
The parameter $\hbar\Omega/E_R$ can be fixed near unity if the Raman laser beams have intensities on the order of
1 mW/cm$^2$ (the intensity of one beam must be more than about ten times the other to satisfy the $\Omega\ll\Delta$
condition) and are detuned from the $^{87}$Rb~$D_2$ line by about $\delta_r\approx500$~GHz.

In cold atom optics, metal-insulator transitions can be detected by measuring the momentum distribution
of the expanding atomic clouds. The Bragg peaks in the metallic phase (as shown in Fig.~\ref{fig4}), reflecting the quasiperiodic order, become flattened at the onset to localization.
It should be noted that in the non-Abelian case, the momentum distribution shows additional satellite peaks around the Bloch vectors.

In this paper, we have identified
means for driving a localization transition in two-dimensional
optical lattices, and have contrasted the effects of different gauge symmetries.
The $U(2)$ gauge system clearly offers a much richer environment than does $U(1)$.
For example, using polarized fermions such as $^{40}K$
and with the Fermi energy in the central band, one may observe the complete defiance of localization.
For a BEC loaded in the lattice,
smooth changes in the ground state energy for the Abelian system will
be replaced by extremely sharp oscillations as the transverse component of the momentum is tuned (see Fig.~\ref{fig1}).
In addition to controlling the nature of the quantum phases
by the ratio of the tunneling in two directions, the transport properties can also be controlled
by the momentum. Such novelties will broaden the possibilities of using ultracold gases
in exploring new horizons in physics.



\begin{references}
\bibitem{foot}  The non-Abelian case with $\alpha_2=0$
provides some insight into this three-fold degenerate point. It is easy to see that
$E=2\Lambda \cos(k_y)$ is always a solution of the TBM corresponding to $\alpha_2=0$.
The $\alpha_2=\frac{1}{2}$ breaks up the curve $E=2\cos(k_y)$ and the three fold degeneracy
is a consequence of the fact
that the $E(k_y,\Lambda)=0$ solution of the $\alpha_2=0$ and $\alpha_2=\frac{1}{2}$ are identical.
\end{references}
\end{document}